\documentclass[final,pdftex,reqno,a4paper,12pt]{myectaart}

\RequirePackage[OT1]{fontenc}
\RequirePackage{amsthm}
\RequirePackage[cmex10]{amsmath}
\RequirePackage{natbib}
\RequirePackage[colorlinks,citecolor=blue,urlcolor=blue]{hyperref}
\RequirePackage{hypernat}

\usepackage{a4wide,graphicx,float,amssymb} 

\startlocaldefs
\numberwithin{equation}{section}
\theoremstyle{plain}

\endlocaldefs

\newtheorem{theorem}{Theorem}

\newtheorem{propos}[theorem]{Proposition}

\RequirePackage{ae,fancyvrb}
\DefineVerbatimEnvironment{Sinput}{Verbatim}{fontshape=sl}
\DefineVerbatimEnvironment{Soutput}{Verbatim}{}
\DefineVerbatimEnvironment{Scode}{Verbatim}{fontshape=sl}

\setkeys{Gin}{width=0.9\textwidth}

\begin{document} 

\begin{frontmatter}
\title{\bf On moment indeterminacy of the Benini income distribution %\protect\thanksref{T1}
}
\runtitle{Moment indeterminacy of the Benini distribution}

%\thankstext{T1}{Version: \today.}

\begin{aug}
\author{\fnms{Christian} \snm{Kleiber}%\thanksref{t1}
\ead[label=e1]{christian.kleiber@unibas.ch}}

%\thankstext{t1}{Version: \today.}

\runauthor{C. Kleiber}

%\thankstext{t1}{Some comment}

\address{Version: April 29, 2013. \ Correspondence to: Christian Kleiber, Faculty of Business and Economics, %Quantitative Methods Unit, 
Universit\"at Basel, Peter Merian-Weg 6, 4002 Basel, Switzerland.
\printead{e1}}
\end{aug}

\begin{abstract}
The Benini distribution is a lognormal-like distribution generalizing the Pareto distribution. Like the Pareto and the lognormal distributions it was originally proposed for modeling economic size distributions, notably the size distribution of personal income. This paper explores a probabilistic property of the Benini distribution, showing that it is not determined by the sequence of its moments  although all the moments are finite. It also provides explicit examples of distributions possessing the same set of moments. Related distributions are briefly explored.

\smallskip
\noindent
{\sc Keywords:} Benini distribution, characterization of distributions, income distribution, moment problem, statistical distributions, Stieltjes class.

\smallskip
\noindent
{\sc AMS 2010 Mathematics Subject Classification:} Primary 60E05, Secondary 62E10, 44A60. 

\smallskip
\noindent
{\sc JEL Classification:} C46, C02. 
\end{abstract}

%\begin{keyword}
%\kwd{Generalized error distribution} 
%\kwd{generalized lognormal distribution, lognormal distribution, moment problem, size distribution, %statistical distributions, 
%Stieltjes class, volatility model}
%\end{keyword}

\end{frontmatter}

\section{Introduction} \label{intro}

In the late 19th century, the eminent Italian economist Vilfredo Pareto observed that empirical income distributions are well described by a straight line on a doubly logarithmic plot \citep{mom:Pareto:1895, mom:Pareto:1896,mom:Pareto:1897}. Specifically, with $\overline{F} = 1 - F$ denoting the survival function of an income distribution with c.d.f. $F$, Pareto observed that, to a good degree of approximation,

\begin{equation}\label{paretocdf}
\ln \overline{F}(x) ~=~ a_0 - a_1 \ln x .
\end{equation}

\noindent
The distribution implied by this equation is called the Pareto distribution. 

Not much later, the Italian statistician and demographer Rodolfo Benini found that a second-order polynomial 

\begin{equation}\label{beninicdf}
\ln \overline{F}(x) ~=~ a_0 - a_1 \ln x - a_2 (\ln x)^2 
\end{equation}

\noindent
sometimes provides a markedly better fit \citep{mom:Benini:1905,mom:Benini:1906}. The distribution implied by this equation is called the Benini distribution.

The present paper is concerned with a probabilistic property of the Benini distribution, namely whether it is possible to characterize this distribution in terms of its moments. The moment problem asks, for a given distribution $F$ with finite moments \ $\mu_k ~\equiv~ \mathbb{E}[X^k] ~=~ \\ \int ^\infty _{-\infty} x^k \hbox{d}F(x)$ of all orders $k = 1, 2, \dots$, whether or not $F$ is uniquely determined by the sequence of its moments. See, for example, \cite{mom:Shohat+Tamarkin:1950} for analytical or \citet[][Sec.~11]{mom:Stoyanov:2013} for probabilistic aspects of the moment problem. If a distribution is uniquely determined by the sequence of its moments it is called moment-determinate, otherwise it is called moment-indeterminate. Cases where the support of the distribution is the positive half-axis $\mathbb{R}^+ = [0, \infty)$ or an unbounded subset thereof are called Stieltjes-type moment problems. The Benini distribution thus poses a Stieltjes-type moment problem. It is shown below that the Benini moment problem is indeterminate. Drawing on a classical example going back to \cite{mom:Stieltjes:1894} explicit examples of distributions possessing the same set of moments are constructed. Certain generalizations of the Benini distribution are briefly explored, all of which are moment-indeterminate.

\section{The Benini distribution}\label{sec:benini}

Pareto's observation (\ref{paretocdf}) leads to a distribution of the form

\[
F(x) ~=~ 1 - \left( \frac{x}{\sigma} \right)^{-\alpha}, \quad x \geq \sigma > 0,
\]

\noindent
where $\alpha > 0$. Benini's observation (\ref{beninicdf}) leads to a distribution of the form 

\begin{equation} \label{benini3cdf}
F(x) ~=~ 1 - \exp \left\{ - \alpha \ln \frac{x}{\sigma} - \beta \left( \ln \frac{x}{\sigma} \right)^2 \right\} , \quad x \geq \sigma > 0,
\end{equation}

\noindent
where $\alpha , \beta \geq 0$, with $(\alpha, \beta) \neq (0,0)$. Setting $\beta = 0$ gives the Pareto distribution.  

For parsimony, \cite{mom:Benini:1905} often worked with the special case where $\alpha = 0$, i.e. with 

\begin{eqnarray} \label{benini2cdf}
F(x) & = & 1 - \exp \left\{- \beta \left( \ln \frac{x}{\sigma} \right)^2 \right\}   \\
\nonumber & = & 1 - \left( \frac{x}{\sigma}\right)^{-\beta (\ln x -
\ln \sigma)} , \quad x \geq \sigma > 0.
\end{eqnarray}

\smallskip\noindent
Here $\sigma > 0$ is a scale and $\beta > 0$ is a shape parameter. This distribution will be denoted as Ben($\beta,\sigma$). For the purposes of the present paper the scale parameter $\sigma$ is immaterial. The object under study is, therefore, the $\hbox{Ben}(\beta,1) \equiv \hbox{Ben}(\beta)$ distribution with

\begin{equation}\label{benini2cdf}
F(x) ~=~ 1 - \exp \{ -\beta (\ln x)^2  \} , \quad x \geq 1.
\end{equation}

It may be worth noting that the Benini distributions are stochastically ordered w.r.t. $\beta$. Specifically, it follows directly from (\ref{benini2cdf}) that 

\begin{equation}\label{stochdom}
F(x; \beta_1) \ \leq \ F(x; \beta_2) \ \mbox{ for all } x \geq 1 \quad \Longleftrightarrow \quad \beta_1 \ \leq \ \beta_2,
\end{equation}

\noindent
hence $F(x; \beta_1)$ is larger than $F(x; \beta_2)$ under this condition in the sense of the usual stochastic order, often called first-order stochastic dominance in economics.

Noting further that the c.d.f. of a Weibull distribution is $F(x) = 1 - \exp(- x^a)$, $x > 0$, $a > 0$, it follows that eq.~(\ref{benini2cdf}) describes a log-Weibull distribution with $a=2$. The Weibull distribution with $a=2$ is also known (up to scale) as the Rayleigh distribution, especially in physics, and so the Benini distribution may be seen as the log-Rayleigh distribution. It may also be seen as a log-chi distribution with two degrees of freedom (again up to scale); i.e., the logarithm of a Benini random variable follows the distribution of the square root of a chi-square random variable with two degrees of freedom.

The density implied by (\ref{benini2cdf}) is 

\begin{equation}\label{benini2pdf}
f(x) ~=~ \frac {2 \beta \ln x}{x} \ \exp\left\{ - \beta (\ln x)^2 \right\}, \quad x \geq 1,
\end{equation}
and hence is similar to the density of the more familiar lognormal distribution. The lognormal distribution is perhaps the most widely known example of a distribution that is not determined by its moments, although all its moments are finite \citep{mom:Heyde:1963}. The similarity of the lognormal and the Benini densities now suggests that the Benini distribution might also possess this somewhat pathological property. The remainder of the present paper explores this issue. 

\begin{figure}[ht!]
\begin{center}
\includegraphics{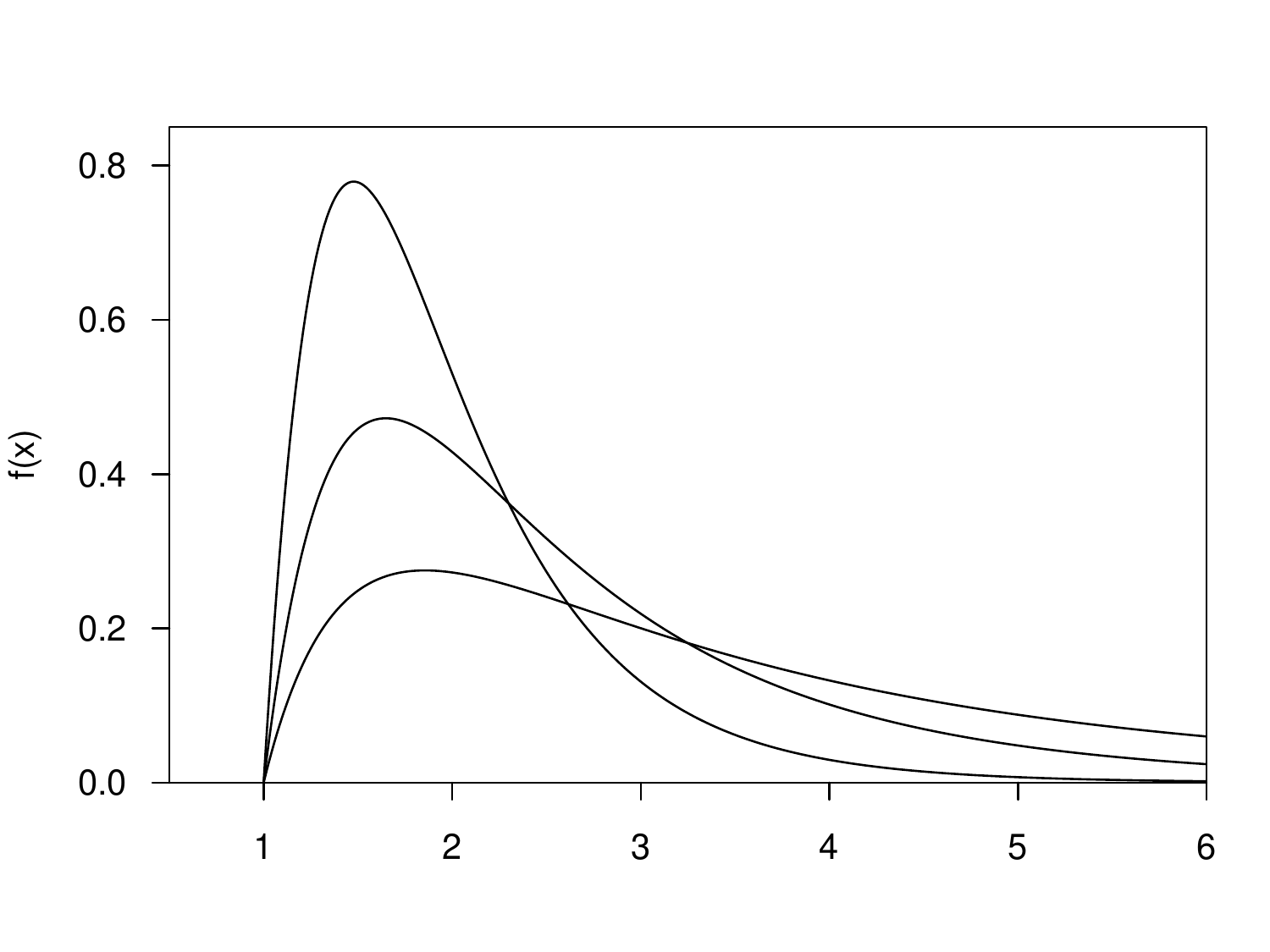}
\end{center}
\caption{\label{fig:fig1} Some Benini densities; $\beta = 2, 1, 0.5$ (from left to right). }
\end{figure}

Figure \ref{fig:fig1} depicts some two-parameter Benini densities, showing that distributions with smaller values of $\beta$ are associated with heavier tails, as indicated by (\ref{stochdom}).

From a modeling point of view, the significance of the Benini distribution lies in the fact that it generalizes the Pareto distribution while itself being `lognormal-like'. It thus enables to discriminate between these two widely used distributions, at least approximately. Further details on the Benini distribution, including an independent rediscovery in actuarial science motivated by failure rate considerations \citep{mom:Shpilberg:1977}, may be found in \citet[][Ch.~7.1]{mom:Kleiber+Kotz:2003}. The appendix of \cite{mom:Kleiber+Kotz:2003} also provides a brief biography of Rodolfo Benini.

\section{The Benini distribution and the moment problem}\label{sec:beninimom}

The following proposition provides two basic properties of the Benini distribution that are relevant in the context of the moment problem.

\begin{propos}\label{proposmom}
\begin{itemize}
\item[(a)] The moments $\mu_k$, $k \in \mathbb{N}$, of the Benini distribution Ben($\beta$) are given by
\begin{eqnarray}
\mu_k ~\equiv~ \mathbb{E}[X^k] &=& 1 + k ~ (2\beta)^{-(1/2)} ~ e^{k^2/(8\beta)} ~ D_{-1}\left( \frac{-k}{\sqrt{2\beta}} \right) \label{mom1} \\
&=& 1 + \frac{k \sqrt{\pi}}{2\sqrt{\beta}} \ e^{k^2/(4\beta)} \ \left\{ 1 + \mathrm{erf} \left( \frac{k}{2\sqrt{\beta}} \right) \right\} . \label{mom2}
\end{eqnarray}
Here, $D_{-1}$ is a parabolic cylinder function and $\mathrm{erf}$ denotes the error function.

\item[(b)] The moment generating function (m.g.f.) of the Benini distribution does not exist.
\end{itemize}
\end{propos}

\medskip
\emph{Proof.} \quad (a) We have
\begin{eqnarray*}
\mu_k ~\equiv~ \mathbb{E}[X^k] 
&=& k \int_0^\infty x^{k-1} \overline{F} (x) \, \hbox{d}x \\
&=& 1 + k \int_1^\infty x^{k-1} \exp\{ -\beta (\ln x)^2 \} \, \hbox{d}x \\ 
&=& 1 + k \int_0^\infty e^{kx - \beta x^2} \, \hbox{d}x \\
&=& 1 + k ~ (2\beta)^{-(1/2)} ~ e^{k^2/(8\beta)} ~ D_{-1}\left( \frac{-k}{\sqrt{2\beta}} \right),
\end{eqnarray*}

\noindent
using \cite{mom:Gradshteyn+Ryzhik:2007}, no. 3.462, eq.~1. This proves (\ref{mom1}). The alternative representation (\ref{mom2}) is established via the relation \citep[][\S~12.7.5]{mom:Olver+Lozier+Boisvert:2010}

\[
D_{-1} (x) ~=~ \sqrt{\frac \pi2 } \ e^{x^2 /4} \ \hbox{erfc} \left( \frac{x}{\sqrt{2}} \right) ,
\]
where $\hbox{erfc}(\cdot)$ is the complementary error function, together with $\hbox{erfc}(x) = 1-\hbox{erf} (x)$ and $\hbox{erf}(-x) = -\hbox{erf} (x)$. 

\medskip
\noindent
(b) The defining integral is  

\[
\mathbb{E}[e^{t X}] = \int_1^\infty e^{tx} \ \frac{2 \beta \ln x}{x} \ \exp\left\{ - \beta (\ln x)^2 \right\} \, \hbox{d}x ~=:~ \int_1^\infty h(x) \, \hbox{d}x .
\]

\noindent
Now the leading term in

\[
\ln h(x) ~=~ tx + \ln (2 \beta \ln x) - \ln x - \beta (\ln x)^2 
\]

\noindent
is the linear term, hence $\mathbb{E}[e^{t X}] = \infty$ for all $t > 0$. \hspace*{\fill} $\square$

\bigskip
The representation (\ref{mom2}) can also be obtained using \textsf{Mathematica} \citep{mom:Wolfram:2013}, version 9.0.1.0.

As an illustration, Table~\ref{tab:tab1} provides the first four moments of selected Benini distributions, namely those from Figure~\ref{fig:fig1}. These moments are rather large, especially for small values of $\beta$.

\begin{table}[ht!]
\caption{Lower-order moments of Benini distributions.}
\begin{center}
\label{tab:tab1}
\begin{tabular}{|p{1.8cm}|r|r|r|r|}
\hline
 & $\mathbb{E}[X]$ & $\mathbb{E}[X^2]$ & $\mathbb{E}[X^3]$ & $\mathbb{E}[X^4]$ \\
 \hline\hline 
$\beta = 2$ & 1.98 & 4.48 & 11.81 & 37.20\\
$\beta = 1$ & 2.73 & 9.88 & 50.59 & 387.19\\
$\beta = 0.5$ & 4.48 & 37.20 & 677.00 & 29888.67\\
\hline
\end{tabular}
\end{center}
\label{default}
\end{table}%

Proposition~\ref{proposmom} showed that the Benini distribution has moments of all orders, but no m.g.f. Distributions possessing these properties are candidates for moment indeterminacy, although these facts alone are not conclusive. Unfortunately, no tractable necessary and sufficient condition  for moment indeterminacy is currently known. 

For exploring determinacy, the Carleman criterion \citep[e.g.][Sec.~11]{mom:Stoyanov:2013} sometimes provides an answer. In a Stieltjes-type problem, the condition

\[
C_S ~:=~\sum_{k=1}^\infty \mu_{k}^{- \frac{1}{2k}} \ = \ \infty
\]
implies that the underlying distribution is characterized by its moments. 

However, Proposition \ref{proposmom} indicates that the moments of the Benini distribution grow rather rapidly. In view of  $\hbox{erf}(x)  \geq 0$, for $x\geq0$, it follows from (\ref{mom2}) that

\[
\mathbb{E}[X^k] ~\geq~ \frac{k \sqrt{\pi}}{2\sqrt{\beta}} \ e^{k^2/(4\beta)} . %= \mathcal{O}(k e^{k^2/4})
\]

\noindent
Using the ratio test this further implies that

\begin{equation}\label{carleman}
C_S ~=~ \sum_{k=1}^\infty \mu_{k}^{- \frac{1}{2k}} \ \leq \ \sum_{k=1}^\infty \left(\frac{2\sqrt{\beta}}{k \sqrt{\pi}} \right) ^{2k} \ e^{-k/(8\beta)} \ < \infty .
\end{equation}
So the Carleman condition cannot establish determinacy here. 

This suggests to explore indeterminacy instead. Indeed, Theorem~\ref{th:mindet} shows that all Benini distributions are moment-indeterminate. Two proofs are given, one utilizing a converse to the Carleman criterion due to \cite{mom:Pakes:2001} and the other utilizing the Krein criterion \citep{mom:Stoyanov:2000,mom:Stoyanov:2013}.

\begin{theorem}\label{th:mindet}
The Benini distribution $\hbox{Ben}(\beta)$ is moment-indeterminate for any $\beta > 0$.
\end{theorem}

\smallskip
\noindent
\emph{Proof 1.} \quad \citet[][Th.~3]{mom:Pakes:2001} showed that if there exists $x_0 \geq 0$ such that $0 < f(x) < \infty$ for $x > x_0$, the condition $C_S < \infty$ together with the convexity of the function $\psi(x) := - \ln f(e^x)$ on the interval $(\ln x_0, \infty)$ implies moment indeterminacy. $C_S < \infty$ was shown in (\ref{carleman}). For the Benini distribution, the function
\[
\psi(x) ~=~ - \ln f(e^x) ~=~ - \ln(2\beta x) + x + \beta x^2 
\]
is easily seen to be convex on the interval $(0, \infty)$ in view of $\beta > 0$. \hfill $\square$

\medskip
\noindent
\emph{Proof 2.} \quad In the case of a Stieltjes-type moment problem, the Krein criterion requires, for a strictly positive density $f$ and some $c > 0$, that the logarithmic integral 

\begin{equation}\label{krein}
K_S [f] ~=~ \int_c ^\infty \frac{- \ln f(x^2)}{1 + x ^2} \ \hbox{d}x 
\end{equation}
is finite. For the Benini distribution this integral is, choosing $c = e$,

\[
K_S[f] ~=~  - \int_e ^\infty \frac { \ln (2 \beta \ln x^2) - \ln x^2 - \beta  ( \ln x^2 )^2 }{ 1 + x^2} \ \hbox{d}x .
\]
This quantity is finite for all $\beta > 0$. \hfill $\square$

\section{A Stieltjes class for the Benini distribution}\label{sec:stieltjes}

The methods used in the proof of Theorem \ref{th:mindet} only establish existence of further distributions possessing the same set of moments as the Benini distribution. It is known from \cite{mom:Berg+Christensen:1981} that if a distribution is moment-indeterminate, then there exist infinitely many continuous and also infinitely many discrete distributions possessing the same moments. It is, therefore, of interest to find explicit examples of such objects.

A Stieltjes class \citep{mom:Stoyanov:2004} corresponding to a moment-indeterminate distribution with density $f$ is a set

\[
\mathcal{S}(f, p) \ = \ \{f_\varepsilon(x) \ | \ f_\varepsilon(x) := f(x)[1 + \varepsilon \ p(x)], \ x \in \hbox{supp}(f) \} , 
\]
where $p$ is a perturbation function satisfying  
$\mathbb{E}[X^k p(X)] = 0$ for all $k = 0,1,2, \dots$. If $-1 \leq p(x) \leq 1$ and $\varepsilon \in [-1, 1]$, then $\mathcal{S}(f, p)$ is called a two-sided Stieltjes class. Counterexamples to moment determinacy in the literature are typically of this type. It is also possible to have one-sided Stieltjes classes, for which $p$ only needs to be bounded from below, and $\varepsilon \geq 0$. The following Theorem provides a one-sided Stieltjes class for the Benini distribution.

\begin{theorem}\label{th:sc} 
The distributions with densities $f_\varepsilon$, $0 \leq \varepsilon \leq 1$, 

\[
f_\varepsilon (x) ~=~ f(x) \left\{ 1 + \varepsilon \ \frac{x \ \exp\{-(x-1)^{1/4} + \beta (\ln x)^2 \}  \sin\{(x-1)^{1/4}\}}{2 C ~ \beta \ln x}  \right\}, \ x \geq 1 ,
\]
all have the same moments as the Benini distribution $\hbox{Ben}(\beta)$ with density $f$. Here $C > 0$ is a normalizing constant defined in the proof.
\end{theorem}

\medskip\noindent
\emph{Proof.} \quad Consider the (unscaled) perturbation
\[
\tilde p(x) ~=~ \frac{x \ \exp\{-(x-1)^{1/4} + \beta (\ln x)^2 \}  \sin\{(x-1)^{1/4}\}}{2 ~ \beta \ln x}, \quad x \geq 1. 
\]
This perturbation has the following properties:

\medskip
\noindent
(P1). $ \lim_{x \to 1^+} \tilde p(x) ~=~ \infty$.

\noindent
(P2). Basic properties of the sine function imply that $\tilde p(x) \geq 0$ on the interval $(1,2]$.

\noindent
(P3). On the interval $[2,\infty)$, the function $\tilde p$ is continuous, with $\tilde p(2) < \infty$ and $\lim_{x \to \infty} \tilde p(x) \\ ~=~ 0$. Hence $\tilde p(x)$ is bounded there.

\smallskip
Let $C = \sup_{x \in [2,\infty)} |\tilde p(x)|$ and set $p(x) = \tilde p(x)/C$. It follows from (P1)--(P3) that $p$ is unbounded from above and bounded from below, specifically $-1 \leq p \leq \infty$. By construction, $f_\varepsilon \geq 0$. The moments of the corresponding random variable $X_\varepsilon$ with density $f_\varepsilon$, $0 \leq \varepsilon \leq 1$, are further given by

\begin{eqnarray*}
\mathbb{E}[X_\varepsilon^k] 
&=& \int_1^\infty x^k f_\varepsilon(x) \, \hbox{d}x \\
&=& \int_1^\infty x^k f(x) \ \{ 1 + \varepsilon \ p(x) \} \, \hbox{d}x \\
&=& \int_1^\infty x^k f(x) \, \hbox{d}x  + \frac \varepsilon C \int_1^\infty x^k \exp\{-(x-1)^{1/4}\}  \sin\{(x-1)^{1/4}\} \, \hbox{d}x \\
&=:& \mathbb{E}[X^k] + J.
\end{eqnarray*}

\noindent
It remains to show that $J = 0$. Now

\begin{eqnarray*}
\lefteqn{\int_1^\infty x^k \exp\{-(x-1)^{1/4}\}  \sin\{(x-1)^{1/4}\} \, \hbox{d}x } \\
&& \qquad \qquad = \int_0^\infty (x+1)^k \exp\{-x^{1/4}\}  \sin\{x^{1/4}\} \, \hbox{d}x \\
&& \qquad \qquad = \sum_{j=0}^k {k \choose j}\int_0^\infty x^{k-j} \exp\{-x^{1/4}\}  \sin\{x^{1/4}\} \, \hbox{d}x \\ 
&& \qquad \qquad = 0
\end{eqnarray*}
in view of 
\begin{equation}\label{stieltjes}
\int_0^\infty x^n \exp\{-x^{1/4}\}  \sin\{x^{1/4}\} \, \hbox{d}x  ~=~ 0
\end{equation} 
for all $n \in \mathbb{N}_0$. In particular, $ \int_0^\infty f_\varepsilon (x) \, \hbox{d}x ~=~ 1$. \hspace*{\fill} $\square$

\bigskip
Note that Theorem~\ref{th:sc} provides a further proof of the moment indeterminacy of the Benini distribution.

Apart from the shifted argument, the perturbation employed here draws on the pioneering work of \cite{mom:Stieltjes:1894}. In modern terminology, Stieltjes showed that the relation (\ref{stieltjes})  leads to a family of distributions whose moments coincide with those of a certain generalized gamma distribution, implying that the latter is moment-indeterminate.  

\cite{mom:Stieltjes:1894} has a further, and more widely known, example of a distribution that is not determined by its moments, the lognormal distribution. The counterexample he provides for that distribution employs the perturbation 

\begin{equation}\label{sheyde}
p(x) ~=~ \sin( 2 \pi \ln x), \quad x > 0 ,
\end{equation}
which was further developed by \cite{mom:Heyde:1963}. It can also lead to a Stieltjes class for the Benini distribution. However, note that in view of the exponential term common to both the lognormal and the Benini densities, the perturbation based on (\ref{sheyde}) only works for small values of $\beta$, otherwise the resulting ratio diverges for $x \to \infty$.  Methods outlined by \cite{mom:Stoyanov+Tolmatz:2005} may help to construct Stieltjes classes based on (\ref{sheyde}) and the lognormal density that cover the entire range of the shape parameter $\beta$, at the price of somewhat greater analytical complexity.

\section{Related distributions}\label{sec:related}

It is natural to augment Pareto's equation (\ref{paretocdf}) by higher-order terms going beyond the second-order term proposed by \cite{mom:Benini:1905}. Not surprisingly, curves of the form

\begin{equation}\label{gbenini}
\ln \overline{F}(x) ~=~ a_0 - a_1 \ln x - a_2 (\ln x)^2 - \ldots - a_k (\ln x)^k
\end{equation}
soon began to appear in the subsequent Italian-language literature on economic statistics; see, e.g., \cite{mom:BrescianiTurroni:1914} and \cite{mom:Mortara:1917} for some early contributions. 
Somewhat later, the Austrian statistician \cite{mom:Winkler:1950} independently also experimented with polynomials in $\ln x$. Specifically, he fitted a quadratic---i.e., the three-parameter Benini distribution (\ref{benini3cdf})---to the U.S. income distribution of 1919.

Dropping a scale parameter, i.e. setting $a_0 = 0$, eq.~(\ref{gbenini}) gives the c.d.f.   

\begin{equation} \label{gbeninicdf}
F(x) ~=~ 1 - \exp \left\{ - \sum_{j=1} ^k a_j \left( \ln  x \right)^j \right\} , \quad x \geq 1,
\end{equation}

\noindent
where $a_1, \dots, a_k \geq 0$, with corresponding density

\begin{equation} \label{gbeninipdf}
f(x) ~=~ \exp \left\{ - \sum_{j=1} ^k a_j \left( \ln x \right)^j \right\} \left\{ \sum_{j=1} ^k j a_j \left( \ln x \right)^{j-1} \right\} \frac{1}{x}, \quad x \geq 1.
\end{equation}

Using the Krein criterion it is not difficult to see that these generalized Benini distributions are moment-indeterminate, provided $(a_2, \dots, a_k) \neq (0, \dots,0)$ as otherwise not all moments exist. 

A further generalization of the Benini distribution proceeds along different lines. 
In section~\ref{sec:benini} it was noted that the Benini distribution may be seen as the log-Rayleigh distribution, up to scale. It is then natural to consider the log-Weibull family, with c.d.f. 

\[
F(x) ~=~ 1 - \exp \{-(\ln x)^a\}, \quad x \geq 1, 
\] 

\noindent
where $a>0$, and corresponding density
\[
f(x) ~=~ \frac{a (\ln x)^{a-1}}{x} \ \exp \{-(\ln x)^a\}, \quad x \geq 1.
\]

Indeed, \citet[][p.~231]{mom:Benini:1905} briefly discusses this model and reports that, for his data, when $a = 2.15$ the fit is superior to the one using model (\ref{benini2cdf}). Again, the Krein criterion may be used to show that the log-Weibull distributions are moment-indeterminate for any $a>0$.

\bibliography{benini}

\begin{thebibliography}{22}
\newcommand{\enquote}[1]{``#1''}
\expandafter\ifx\csname natexlab\endcsname\relax\def\natexlab#1{#1}\fi

\bibitem[\protect\citeauthoryear{Benini}{Benini}{1905}]{mom:Benini:1905}
\textsc{Benini, R.} (1905): \enquote{I diagrammi a scala logaritmica (a
  proposito della graduazione per valore delle successioni ereditarie in
  {I}talia, {F}rancia e {I}nghilterra),} \emph{Giornale degli Economisti, Serie
  II}, 16, 222--231.

\bibitem[\protect\citeauthoryear{Benini}{Benini}{1906}]{mom:Benini:1906}
---\hspace{-.1pt}---\hspace{-.1pt}--- (1906): \emph{Principii di Statistica
  Metodologica}, Torino: Unione Tipografica--Editrice Torinese.

\bibitem[\protect\citeauthoryear{Berg and Christensen}{Berg and
  Christensen}{1981}]{mom:Berg+Christensen:1981}
\textsc{Berg, C. and J.~P.~R. Christensen} (1981): \enquote{Density Questions
  in the Classical Theory of Moments,} \emph{Annales de l'Institut Fourier},
  31, 99--114.

\bibitem[\protect\citeauthoryear{{Bresciani Turroni}}{{Bresciani
  Turroni}}{1914}]{mom:BrescianiTurroni:1914}
\textsc{{Bresciani Turroni}, C.} (1914): \enquote{Osservazioni critiche sul
  ``{M}etodo di {W}olf'' per lo studio della distribuzione dei redditi,}
  \emph{Giornale degli Economisti e Rivista di Statistica, Serie IV}, 25,
  382--394.

\bibitem[\protect\citeauthoryear{Gradshteyn and Ryzhik}{Gradshteyn and
  Ryzhik}{2007}]{mom:Gradshteyn+Ryzhik:2007}
\textsc{Gradshteyn, I.~S. and I.~M. Ryzhik} (2007): \emph{Tables of Integrals,
  Series and Products}, Amsterdam: Academic Press, 7th ed.

\bibitem[\protect\citeauthoryear{Heyde}{Heyde}{1963}]{mom:Heyde:1963}
\textsc{Heyde, C.~C.} (1963): \enquote{On a Property of the Lognormal
  Distribution,} \emph{Journal of the Royal Statistical Society, Series B}, 25,
  392--393.

\bibitem[\protect\citeauthoryear{Kleiber and Kotz}{Kleiber and
  Kotz}{2003}]{mom:Kleiber+Kotz:2003}
\textsc{Kleiber, C. and S.~Kotz} (2003): \emph{Statistical Size Distributions
  in Economics and Actuarial Sciences}, Hoboken, NJ: John Wiley \& Sons.

\bibitem[\protect\citeauthoryear{Mortara}{Mortara}{1917}]{mom:Mortara:1917}
\textsc{Mortara, G.} (1917): \emph{Elementi di Statistica}, Rome: Athenaeum.

\bibitem[\protect\citeauthoryear{Olver, Lozier, Boisvert, and Clark}{Olver
  et~al.}{2010}]{mom:Olver+Lozier+Boisvert:2010}
\textsc{Olver, F. W.~J., D.~W. Lozier, R.~F. Boisvert, and C.~W. Clark}, eds.
  (2010): \emph{{NIST} Handbook of Mathematical Functions}, Cambridge:
  Cambridge University Press.

\bibitem[\protect\citeauthoryear{Pakes}{Pakes}{2001}]{mom:Pakes:2001}
\textsc{Pakes, A.~G.} (2001): \enquote{Remarks on Converse {C}arleman and
  {K}rein Criteria for the Classical Moment Problem,} \emph{Journal of the
  Australian Mathematical Society}, 71, 81--104.

\bibitem[\protect\citeauthoryear{Pareto}{Pareto}{1895}]{mom:Pareto:1895}
\textsc{Pareto, V.} (1895): \enquote{La legge della domanda,} \emph{Giornale
  degli Economisti}, 10, 59--68, {E}nglish translation in {\it Rivista di
  Politica Economica}, 87 (1997), 691--700.

\bibitem[\protect\citeauthoryear{Pareto}{Pareto}{1896}]{mom:Pareto:1896}
---\hspace{-.1pt}---\hspace{-.1pt}--- (1896): \enquote{La courbe de la
  r\'{e}partition de la richesse,} in \emph{Recueil publi\'{e} par la
  Facult\'{e} de Droit \`{a} l'occasion de l'Exposition Nationale Suisse,
  Gen\`{e}ve 1896}, Lausanne: Ch. Viret-Genton, 373--387, {E}nglish translation
  in {\it Rivista di Politica Economica}, 87 (1997), 645--660.

\bibitem[\protect\citeauthoryear{Pareto}{Pareto}{1897}]{mom:Pareto:1897}
---\hspace{-.1pt}---\hspace{-.1pt}--- (1897): \emph{Cours d'\'{e}conomie
  politique}, Lausanne: Ed. Rouge.

\bibitem[\protect\citeauthoryear{Shohat and Tamarkin}{Shohat and
  Tamarkin}{1950}]{mom:Shohat+Tamarkin:1950}
\textsc{Shohat, J.~A. and J.~D. Tamarkin} (1950): \emph{The Problem of
  Moments}, Providence, RI: American Mathematical Society, revised ed.

\bibitem[\protect\citeauthoryear{Shpilberg}{Shpilberg}{1977}]{mom:Shpilberg:1977}
\textsc{Shpilberg, D.~C.} (1977): \enquote{The Probability Distribution of Fire
  Loss Amount,} \emph{Journal of Risk and Insurance}, 44, 103--115.

\bibitem[\protect\citeauthoryear{Stieltjes}{Stieltjes}{1894/1895}]{mom:Stieltjes:1894}
\textsc{Stieltjes, T.~J.} (1894/1895): \enquote{Recherches sur les fractions
  continues,} \emph{Annales de la Facult\'{e} des Sciences de Toulouse}, 8/9,
  1--122, 1--47.

\bibitem[\protect\citeauthoryear{Stoyanov}{Stoyanov}{2000}]{mom:Stoyanov:2000}
\textsc{Stoyanov, J.} (2000): \enquote{Krein Condition in Probabilistic Moment
  Problems,} \emph{Bernoulli}, 6, 939--949.

\bibitem[\protect\citeauthoryear{Stoyanov}{Stoyanov}{2004}]{mom:Stoyanov:2004}
---\hspace{-.1pt}---\hspace{-.1pt}--- (2004): \enquote{Stieltjes Classes for
  Moment-Indeterminate Probability Distributions,} \emph{Journal of Applied
  Probability}, 41A, 281--294.

\bibitem[\protect\citeauthoryear{Stoyanov}{Stoyanov}{2013}]{mom:Stoyanov:2013}
---\hspace{-.1pt}---\hspace{-.1pt}--- (2013): \emph{Counterexamples in
  Probability}, New York: Dover Publications, 3rd ed.

\bibitem[\protect\citeauthoryear{Stoyanov and Tolmatz}{Stoyanov and
  Tolmatz}{2005}]{mom:Stoyanov+Tolmatz:2005}
\textsc{Stoyanov, J. and L.~Tolmatz} (2005): \enquote{Methods for Constructing
  {S}tieltjes Classes for {M}-Indeterminate Probability Distributions,}
  \emph{Applied Mathematics and Computation}, 165, 669--685.

\bibitem[\protect\citeauthoryear{Winkler}{Winkler}{1950}]{mom:Winkler:1950}
\textsc{Winkler, W.} (1950): \enquote{The Corrected {P}areto Law and its
  Economic Meaning,} \emph{Bulletin of the International Statistical
  Institute}, 32, 441--449.

\bibitem[\protect\citeauthoryear{{Wolfram Research, Inc.}}{{Wolfram Research,
  Inc.}}{2013}]{mom:Wolfram:2013}
\textsc{{Wolfram Research, Inc.}} (2013): \emph{\textsf{Mathematica}, Version
  9.0.1.0}, Champaign, IL.

\end{thebibliography}

\end{document}